\documentclass{aa}

\usepackage[T1]{fontenc}
\usepackage{amsmath}
\usepackage{verbatim}
\usepackage{amssymb}	
\usepackage{natbib}
\bibpunct{(}{)}{;}{a}{}{,} 

\usepackage{url}
\usepackage{graphicx}
\usepackage{txfonts}
\usepackage{xcolor}

\begin{document}

   \title{A Pearl in the Shell: an ultra-compact dwarf 
   within the tidal debris surrounding
  spiral galaxy
   NGC~7531}
   \titlerunning{UCD in the shell around NGC~7531}


   \author{David Mart{\'\i}nez-Delgado
          \inst{1,2}\fnmsep\thanks{ARAID Fellow, dmartinez@cefca.es}
          \and 
          Aaron J.~Romanowsky\inst{3,4}          
          \and 
           Yimeng Tang\inst{4}
         \and 
         Joanna D. Sakowska\inst{5,6}
         \and
          Denis Erkal\inst{6}
          \and
         Juan Mir\'o-Carretero\inst{7,8}
         \and
            Giuseppe Donatiello  \inst{9}
          \and
          Sepideh Eskandarlou\inst{1}
          \and
          Mark Hanson\inst{10}
          \and
          Dustin Lang\inst{11}
          }

   \institute{ Centro de Estudios de F\'isica del Cosmos de Arag\'on (CEFCA), Unidad Asociada al CSIC, Plaza San Juan 1, 44001 Teruel, Spain
   \and
ARAID Foundation, Avda. de Ranillas, 1-D, E-50018 Zaragoza, Spain
    \and
Department of Physics \& Astronomy, San Jos\'e State University, One Washington Square, San Jose, CA 95192, USA
 \and
Department of Astronomy \& Astrophysics, University of California, Santa Cruz, CA 95064, USA
 \and
 Instituto de Astrofísica de Andalucía, CSIC, Glorieta de la Astronom\'\i a,  E-18080 Granada, Spain
 \and
School of Mathematics and Physics, University of Surrey, Guildford, GU2 7XH, UK
     \and
    Departamento de F{\'\i}sica de la Tierra y Astrof{\'\i}sica, Universidad Complutense de Madrid, E-28040 Madrid, Spain
    \and
    Leiden Observatory, Leiden University, Gorlaeus Building at Einsteinweg 55, NL-2333 CC Leiden, The Netherlands
  \and
    UAI - Unione Astrofili Italiani /P.I. Sezione Nazionale di Ricerca Profondo Cielo, 72024 Oria, Italy 
    \and
    Doc Greiner Research Observatory-Rancho Hidalgo, Animas, New Mexico, USA 
    \and
    Perimeter Institute for Theoretical Physics,
    31 Caroline St N, Waterloo, Canada}
   \date{Received XXXX; accepted XXXX}

 
  \abstract
    {Stellar substructures within tidal debris preserve information about their progenitor galaxies' properties, offering insights into hierarchical mass assembly processes.}
   {We examine a compact stellar system (CSS) around the nearby spiral galaxy NGC~7531, including the shell-like tidal debris it is embedded within. Our goals are to determine the nature of the CSS, reconstruct its accretion history, and understand how the large, diffuse shell-like structure formed.}
  {We present photometric measurements of the shell-like debris and CSS using DESI Legacy Imaging Survey (LS) data. We obtained Keck/LRIS spectroscopic data for the CSS to confirm its association with NGC~7531 and to
  derive its star formation history through spectral energy distribution fitting. Deep ($\sim$27.9 mag/arcsec$^{2}$) amateur telescope images of NGC 7531 enabled complete characterization of the tidal debris structure. We constructed tailored N-body simulations to reproduce the observed morphology.}
   {We confirm the CSS is associated with NGC~7531. We rename it as NGC 7531-UCD1, since our estimates for its stellar mass ($3.7_{-0.7}^{+1.0}\times 10^6$ $\mathrm{M}_\odot$), half-light radius ($R_{h} = 0.13 \pm 0.05$ arcsec) and extended star formation history place it in the ultra-compact dwarf galaxy (UCD) category. We find NGC 7531-UCD1 experienced a star formation burst $\sim$~1 Gyr ago.  NGC 7531-UCD1 was likely a nuclear star cluster (NSC) that experienced (and may still be experiencing) tidal stripping which transformed it into a UCD -- which is further supported by the presence of tidal tails. We quantify the shell-like debris' mass as $M_\star\sim 3$--$11\times 10^8 M_\odot$, implying a merger mass ratio of $\sim$ 300:1 to 10:1. Our amateur telescope follow-up images confirm new pieces of tidal debris, previously unclear in the DESI LS images. N-body simulations successfully reproduce these tidal features, requiring a near radial orbit of the progenitor dwarf galaxy with two pericentric passages. The timing of the first pericentre passage coincides with the measured star formation enhancement $\sim$~1 Gyr ago.
   }
   {Our findings agree with theoretical predictions about the NSC to UCD formation pathway via tidal stripping, and further confirm the presence of these objects outside of our Milky Way.}


   \maketitle
%
\section{Introduction}\label{sec:intro}

\begin{figure*}
\centering
  \includegraphics[width=0.95\textwidth]{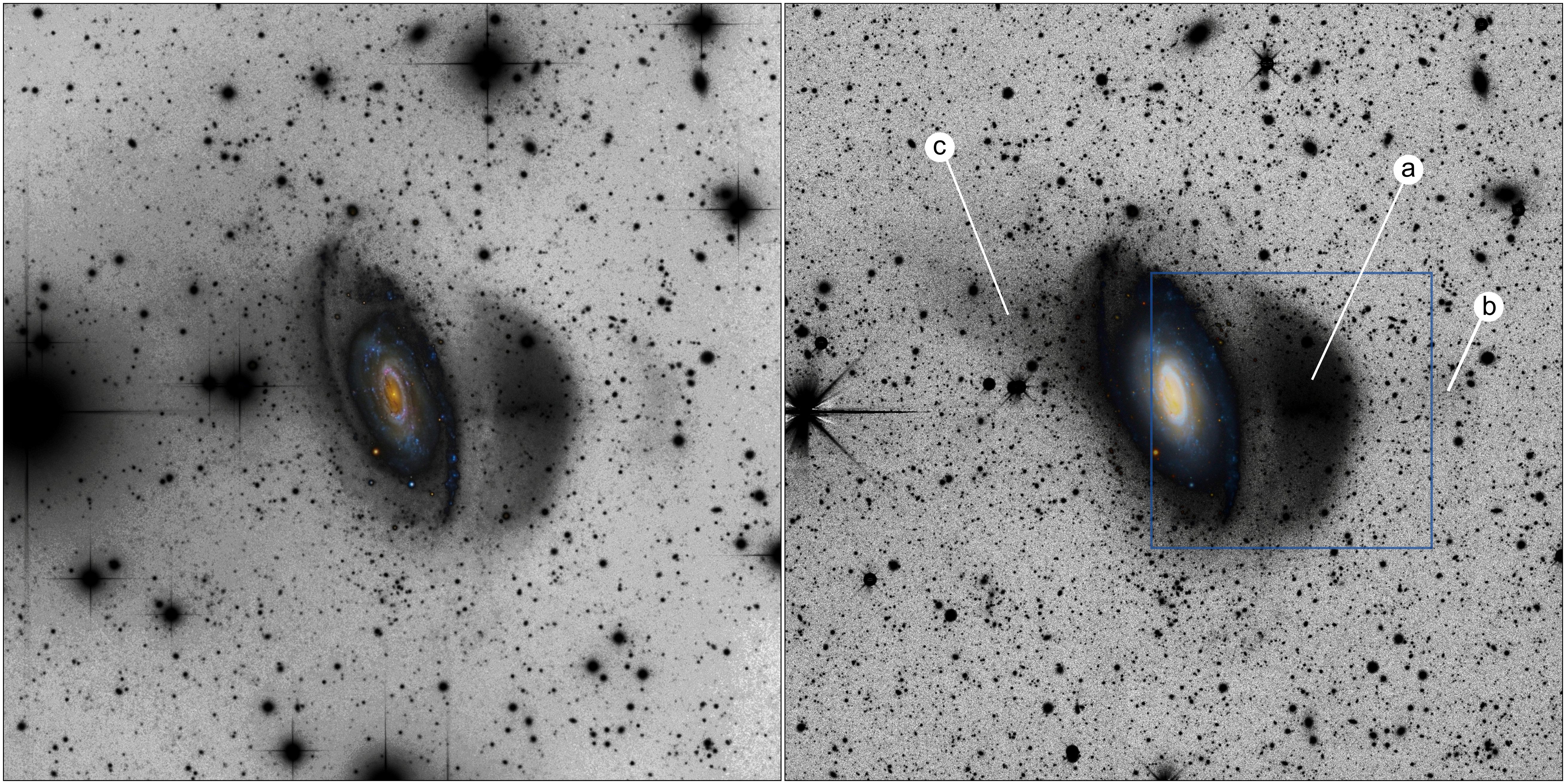}
  \caption{Amateur image of NGC~7531 ({\it left}). DESI Legacy Imaging Survey image of NGC~7531 ({\it right}). Sky-subtracted image of NGC 7531 as processed by Gnuastro's {\it NoiseChisel} program was used as a basis for photometry measurements.
  Features are labelled: (a): main shell; (b): faint outer shell; c: counter plume.  }  
  \label{fig-NGC7531}
\end{figure*}

The outer regions of galaxies, known as galactic halos, contain numerous accreted globular clusters (GCs) and stellar substructures formed when satellite galaxies merge with their hosts and break apart into debris (e.g., \citealt{Searle1978}). Galactic halos, therefore, preserve a rich record of their mass assembly history. For the Milky Way (MW), the advent of the {\it Gaia} satellite (\citealt{prusti2016gaia}) has made it possible to work out individual accretion events in the Galactic halo using multi-dimensional phase-space information of position, velocity, and chemical abundances (e.g., \citealt{Massari2019,Bonaca2024}). For external galaxies, however, we have obtained phase-space information for only a handful of systems, and it has been limited to tracing the most recent, ongoing accretion events (e.g., M31, M87;
\citealt{Romanowsky2012,Dey2023}).

An important part of this picture is that many of the disrupted satellite dwarf galaxies once harbored nuclear star clusters within their cores (NSCs; see \citealt{Neumayer2020} for a modern review). Thanks to their high densities and stellar masses ($M_\star \sim 10^{5-8} M_\odot$), NSCs tend to survive the tidal disruption of their host galaxy and morphologically evolve into a different object class, such as GCs (e.g. \citealt{Bassino1994}) or ultra-compact dwarf galaxies (UCDs; e.g.  \citealt{bekki2003, Pfeffer2013}), the latter of which have relatively large sizes ($10<r_{h}<100$ pc) and stellar masses ($M_\star \sim 10^{6-8} M_\odot$). GCs and UCDs associated with a tidally disrupted dwarf have been directly observed (e.g. \citealt{Foster2014, Jennings2015, Wang2023,Southon2025}), although a statistically large sample of NSC/GC/UCD objects is required to further understand this evolutionary pathway. 

To this end, we focus on the nearby spiral galaxy NGC~7531. NGC~7531 hosts an enormous cloud of shell-like tidal stellar debris, first reported by \cite{Buta1987} from
an inspection of photographic plates. Since then, deeper images of NGC~7531 were provided by the Stellar Tidal Stream survey: a professional--amateur (ProAm) collaboration involving astrophotographers worldwide operating robotic telescopes to obtain unprecedentedly deep images of spiral galaxies (see \citealt{MD2008} for survey description and \citealt{MD2025} for a compilation of results). The resulting luminance-filter images clearly resolved the shape, extension and even the small-scale substructures embedded within the cloud (\citealt{MD2010}). 

In this paper, we re-visit the galaxy in order to understand the nature of the stellar debris and the visually identified compact stellar system (CSS) encompassed within it. In Section \ref{sec:observations} we overview the data used, including the new amateur telescope image of NGC~7531 and the spectroscopic data taken of the CSS object with the LRIS spectrograph (\citealt{LRIS}). In Section \ref{sec:stream}, we perform photometry of the stellar shell and measure its luminosity and stellar mass. In Section \ref{sec:4}, we present the photometry  and spectroscopy of the CSS, including a derivation of its star formation history (SFH) via SED fitting. Finally, in Section \ref{sec:discussion}, we discuss our findings and reproduce the tidal debris and the most-likely recent merger history of NGC~7531 using N-body simulations. In Section \ref{sec:conclusions} we present our conclusions on the nature of the debris and compact object.

\section{Observations and data reduction}\label{sec:observations}




The CSS was identified as a candidate for a massive star cluster using 
{\it Gaia} data and following methods from \cite{Voggel2020}. Star clusters with small enough distances and large enough sizes
can be identified as more extended than point sources through {\it Gaia} excess factors. The CSS has BP/RP=1.83 excess factor and Astrometric Excess Noise=5.6 in Gaia DR3 (see \citealt{gaiadr3} for a description of these parameters) --  values that are somewhat elevated compared to a typical foreground stellar source with a {\it Gaia} magnitude of $G=20.9$ (see \citealt{Hughes2021,Forbes2022}). The object is too faint for a high-quality proper motion or parallax measurement that could identify it as a clear foreground star. Follow-up spectroscopy is therefore required for confirmation as a star cluster associated with NGC~7531.

Below we present imaging data from the DESI Imaging Legacy survey (\S~\ref{sec:data1}) and from an amateur telescope (\S~\ref{sec:data2}), along with spectroscopy from the Keck telescope (\S~\ref{sec:data3}).

\subsection{DESI Legacy surveys imaging data}
\label{sec:data1} 

An image cutout from the DESI Legacy Imaging Survey (DESI LS; \citealt{2019AJ....157..168D}) DR 10, with 20$\times$15 arcmin size and NGC~7531 in the centre, was used as the basis for the photometric analysis (see Figure~\ref{fig-NGC7531}). It was obtained using a modified version of the DESI LS reduction pipeline {\tt Legacypipe}, which alters the way the image backgrounds (``sky models'') are computed. By default, the pipeline uses a flexible spline sky model which can over-subtract the outskirts of large galaxies.  Instead, we subtracted the sky background from each CCD using a custom algorithm, which preserves the low-surface-brightness galactic features of interest. We first minimized the relative background levels between the overlapping CCDs in each band, and then, after detecting and masking sources as well as {\it Gaia} stars, we subtracted the sigma-clipped median in the outer half of the image cutout (see \citealt{MD2021} for details). 

The depth of the DES LS image in each band has been determined by calculating the surface brightness limit following the standard method of \cite{Roman2020}, i.e. the surface brightness corresponding to $3\sigma$ of the signal in the non-detection areas of the image for a 100 arcsec$^2$ aperture.  This yields 29.33, 28.95 and 27.76 mag/arcsec$^{-2}$ for the $g$, the $r$ and the $z$ passbands, respectively.

\subsection{Amateur telescope data}
\label{sec:data2} 

We collected deep imaging of  NGC~7531 at the Mount Lemmon Sky Center (Steward Observatory, University of Arizona) with an 60 cm aperture $f/6.7$ Planewave CDK telescope. We used a SBIG STX 16803  CCD camera was that provided a pixel scale of 0.47$\arcsec$\,pixel$^{-1}$ over a $36.9\arcmin \times 36.9\arcmin$ field of view. We obtained a set of 126 individual 900 second  images with an Astrodon Gen2 Tru-Balance E-series luminance filter over several nights between 2022 September 16th and 2022 September 23th. The photometric zeropoint of the luminance-filter data was obtained by using the DESI LS images as a reference. For this purpose, we combined the DESI LS $g$- and $r$- filters to achieve a similar wavelength coverage. Aperture photometry was conducted on stars selected from {\it Gaia} using {\it GNU Astronomy Utilities} (Gnuastro\footnote{\url{http://www.gnu.org/software/gnuastro}}, \citealt{Akhlaghi2015,Akhlaghi2019})\ Specifically, we used Gnuastro's {\tt MakeProfiles} (for building the apertures), {\tt MakeCatalog}\footnote{\url{https://www.gnu.org/software//gnuastro/manual/html_node/MakeCatalog.html}} for photometry and {\it Query} (for access to the {\it Gaia} database). {\it Gaia} DR3 was used for selecting stars with significant ($\geq3\sigma$) parallax.

To account for the different point spread functions (PSFs) of the DESI LS image and the CDK24 image, we used apertures with radii from $1\arcsec$ to $4\arcsec$ (with a step of $0.25\arcsec$).
The $3.75\arcsec$ aperture had the least scatter in its measured zero-point ($28.65\pm0.04$).
The $3\sigma$ surface brightness limit over $100$ arcsec$^2$ is $27.9$ mag arcsec$^{-2}$, measured using Gnuastro's {\tt NoiseChisel}
and {\tt MakeCatalog}.

\subsection{Keck LRIS spectroscopy}
\label{sec:data3} 

\begin{figure}
  \includegraphics[width=1.00\columnwidth]{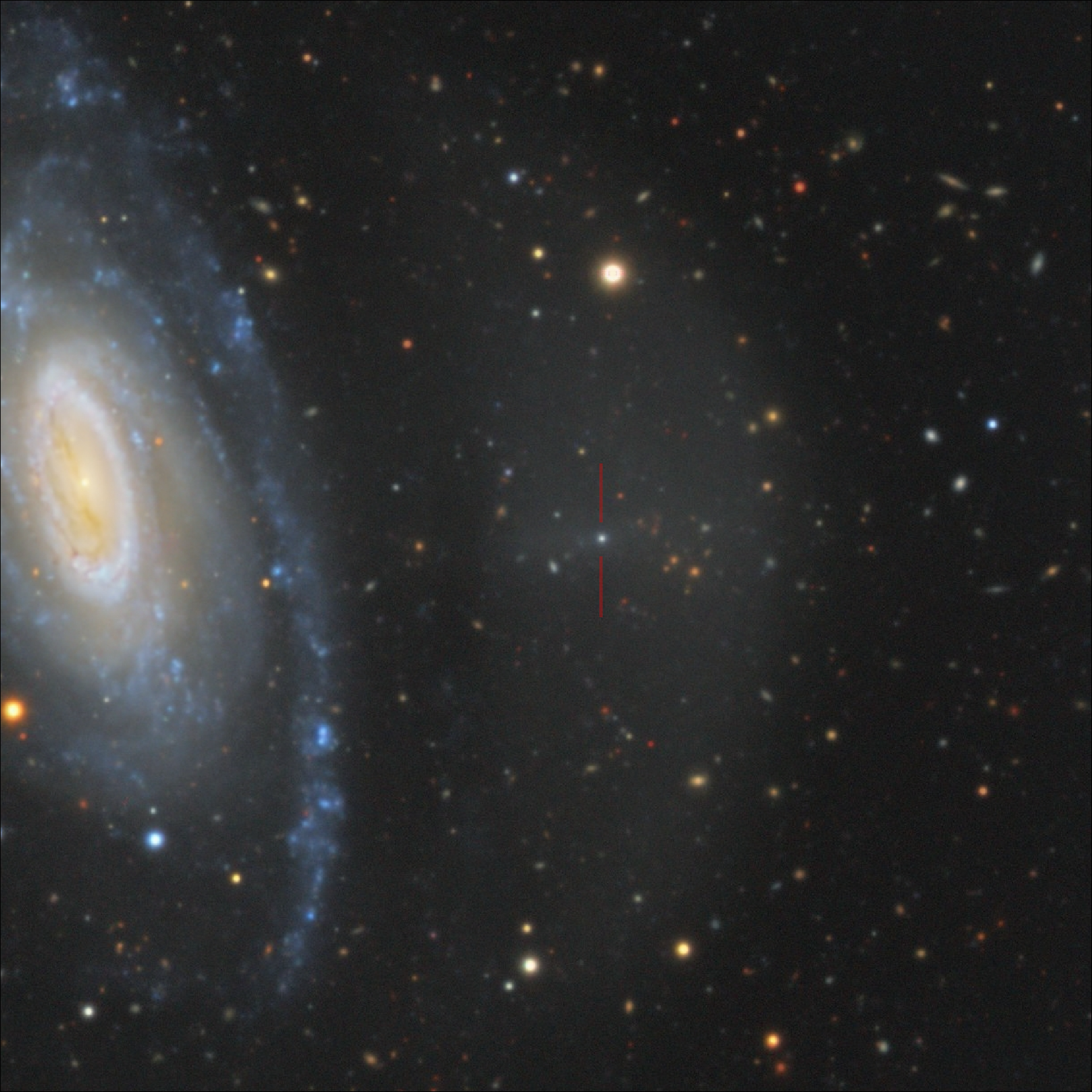}
  \caption{Identification chart of the compact stellar system potentially embedded within the giant tidal debris cloud in the halo of NGC 7531. The background image used is from the DESI Legacy Imaging Survey.}  
  \label{fig-cluster}
\end{figure}

Spectroscopy of the compact object (Figure~\ref{fig-cluster}) was obtained on 2022-09-30 at the Keck Observatory, using the LRIS spectrograph (\citealt{LRIS}). The seeing during the observations was $\sim$0.5--0.7$\arcsec$. The target was observed using a $1.5\arcsec$-wide slit with a position angle of 0 degrees.
    We used the 300/5000 grism in the blue, the D680 dichroic and the 400/8500 grating in the red, targeting a central wavelength of 8500 \AA. 
    Two exposures of 300s each were obtained. The blue side data were reduced using the \texttt{pypeit} reduction pipeline (\citealt{pypeit}) with the nominal settings for the LRIS blue side. The red side was reduced using the settings for the Mark 4 red detector. 
    The \texttt{pypeit} pipeline performs a first order wavelength calibration using arc lamps, applies a flexure correction 
    to each frame based on skyline positions and applies a heliocentric correction to the velocities; this was $-14.87$ km s$^{-1}$ for this object.
    Based on a direct comparison of the wavelength solution to the UVES sky spectrum (\citealt{uves}), we estimate a conservative uncertainty
    in the wavelength solution to be $\sim$2.5 \AA, or $\sim$ 187  km s$^{-1}$, driven primarily by the low resolution of the grism.

    The velocity of the source was fitted in the blue using the Penalized Pixel Fitting (pPXF, \citealt{Cappellari2004, Cappellari2017}).
    The blue side was chosen due to the presence of strong Calcium H+K lines, among other strong absorption features in the blue; the Ca~II triplet was detected in the red, but few other features were present.
    The fit was performed with outlier clipping, and an 8th degree polynomial was fitted to the instrumental response. 
    The $\chi^2/\rm DOF$ of the best fit was 0.47, and used a combination of four templates. 
    The best-fit velocity from  \texttt{pPXF} is 1754 km s$^{-1}$, with a formal uncertainty of 27 km s$^{-1}$. 
    We quote a total velocity uncertainty of 189 km s$^{-1}$ after combining the \texttt{pPXF} and wavelength solution uncertainties in quadrature. 
    The spectrum will be presented later in the paper in the context of stellar population modeling.

\section{The stellar shell of NGC 7531}
\label{sec:stream}

\subsection{Photometry}
\label{sec:photometry1} 

The photometry of the stellar stream around NGC 7531 in the $g$, $r$ and $z$-bands was derived with Gnuastro using the resulting coadded image cutout of this galaxy from the DESI LS data (see Sec.~\ref{sec:data1}). The measurements were carried out with Gnuastro's {\sc MakeCatalog} on the basis of the sky-subtracted image generated by Gnuastro's {\sc NoiseChisel}.

\begin{figure}
  \includegraphics[width=0.95\columnwidth]{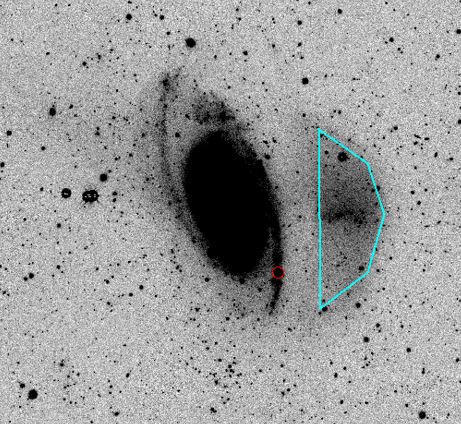}
  \caption{Photometric measurement method for the shell around NGC 7531. The polygonal aperture ({\it blue}) indicates the part of the image where the magnitude, surface brightness and colour of the shell were measured. For comparison, the surface brightness and colours were measured in a circular aperture ({\it red}) placed on the NGC 7531 spiral arm. }  
  \label{fig-photometry}
\end{figure}

The photometric measurements were carried out in a hand-drawn polygonal aperture, covering the clearly detected part of the shell (see Figure \ref{fig-photometry}). Regions where the stream surface brightness could be contaminated by the outer disk of NGC~7531 were avoided. 
The full height and width of the aperture are approximately $4\times1.5$~arcmin ($26\times 9.5$~kpc).
After subtracting the sky from the input images\footnote{Gnuastro applies tessellation to the image for the detection step, so that the sky estimation and subtraction is done on a local tile basis (40$\times$40 pixels tiles were used for the 2290$\times$2290 pixel NGC~7531 image)}, the first step is to perform the detection of the signal. Then all foreground and background objects, identified as {\it clumps} in Gnuastro's {\sc Segment} program, were masked and the photometric parameters (apparent magnitude, surface brightness and colors) were measured in the non-masked part of the polygonal aperture. For comparison, the surface brightness and the color have also been measured in an aperture placed on the galaxy (see red circle in Figure \ref{fig-photometry}). 

The measurement results for the surface brightness and color along with the corresponding uncertainty for the shell around NGC 7531 are given in Table~\ref{tab-photometry}, as computed by Gnuastro's {\sc MakeCatalog}.  The uncertainty in the measurement of the magnitude was calculated with the expression\footnote{\url{https://www.gnu.org/software/gnuastro/manual/html_node/Magnitude-measurement-error-of-each-detection.html}}   
\(M_{error} = 2.5\, /\, S/N\, \ln(10)\).   
We note that with its large size and low surface brightness
($\langle\mu_g\rangle = 25.4$ mag arcsec$^{-2}$), the shell could be classified as an
ultra-diffuse galaxy, belonging to the ``misfit'' variety of tidally distorted systems
\citep{Zaritsky2025}.

\begin{table*}
\centering                          
{\small
\caption{Surface brightnesses and colors for the shell-like debris feature around NGC 7531 (top row) and for an aperture placed on the arm of NGC 7531 (bottom row).}
\label{tab-photometry}
\renewcommand{\arraystretch}{1.5}
\begin{tabular}{l c c c c c c}       
\hline\hline                 
 & <$\mu_{g}$> & <$\mu_{r}$> & <$\mu_{z}$> & <\textit{g}$-$\textit{r}$>$ & <\textit{g}$-$\textit{z}$>$ & <\textit{r}$-$\textit{z}$>$ \\

       & [mag arcsec$^{-2}$] & [mag arcsec$^{-2}$] & [mag arcsec$^{-2}$] & [mag] & [mag] & [mag] \\
\hline                        
 Shell & 25.43 $\pm$ 0.0019 & 24.85 $\pm$ 0.0017 & 24.50 $\pm$ 0.0012 & 0.57 $\pm$ 0.003 & 0.90 $\pm$ 0.002 & 0.34 $\pm$ 0.002 \\
NGC~7531  & 23.79 $\pm$ 0.008 & 23.60 $\pm$ 0.007 & 23.53 $\pm$ 0.007 & 0.18 $\pm$ 0.009 & 0.24 $\pm$ 0.009 & 0.06 $\pm$ 0.009 \\ 
\hline                                   
\end{tabular}
}
\end{table*}

\subsection{Luminosity and Stellar Mass}
\label{sec:pholometry2} 

We next estimate the stellar mass of the shell's progenitor, 
based on the total luminosity of the shell. We are limited by how much of the shell we can actually detect, and there could be regions of the shell that are missed, below the surface brightness limit of the image.
Nevertheless, having a deep image from DESI LS and detecting a sharp boundary of the shell gives us confidence that our estimate should be close to the real value. 
The apparent magnitude was measured from a polygonal aperture covering as much of the shell area as possible, once the foreground and background sources were masked, following the same approach as for the surface brightness and colours, described in Section~\ref{sec:photometry1}.  

From the apparent magnitudes, the estimated distance of NGC~7531 ($D=22.2 \pm 4.9$ Mpc) and the Galactic extinction, both obtained from the {\it NASA/IPAC Extragalactic Database}\footnote{\url{https://ned.ipac.caltech.edu/extinction_calculator}}), we derived the absolute magnitudes for the $g$, $r$ and $z$ passbands. We then computed the luminosity of the shell by using the solar absolute magnitudes for the $g$, the $r$ and the $z$ passbands from \citet{willmer2018}. The results are given in Table~\ref{tab-luminositymass}.
We calculated the mass--to--light ratio ($M/L_{\lambda}$) from the three colours measured for the stream (see Section \ref{sec:photometry1}), using the correlations between SDSS {\it ugriz} colors and SDSS/2MASS $M/L$ values given in \cite{Bell2001}. Following this method, we obtained an estimate for the stellar mass of the stream progenitor between $3.35\times 10^8$  M$_{\odot}$ and $1.1\times 10^9$ M$_{\odot}$. The range of the estimated progenitor's mass is quite large (0.8$\times$10$^8$ M$_\odot$) because the distance estimate for NGC 7531 has a large uncertainty (4.9 Mpc), and also, after measuring three colours, we can apply three sets of parameters in the mass estimate correlation for each passband, yielding nine estimates in total. The range of uncertainty in the mass estimation is obtained by taking the overall largest and overall lowest estimates as the range limits. Considering the log stellar mass of NGC~7531 is $10.5\pm0.5$ in Solar units \citep {2019JPhCS1234a2015A}, the (stellar) mass merger ratio is between ${0.0033}$ and ${0.11}$, so the mass ratio between the host galaxy and the shell progenitor dwarf galaxy is roughly between 10/1 and 300/1. This ratio is key to determining the kind of tidal interaction between the galaxies and to characterizing the resulting merger event. 

\begin{table}
\centering                          
{\small
\caption{Apparent magnitude, absolute magnitude and luminosity obtained for the shell feature detected around NGC 7531.}
\label{tab-luminositymass}
\renewcommand{\arraystretch}{2.0}
\begin{tabular}{l c c c}       
\hline\hline                 
 & magnitude & Magnitude & luminosity \\
       & [mag] & [mag] & [L$_{\odot}$]  \\
\hline                        
 $g$ & 15.56 $\pm$ 0.002 & $-16.6^{+0.4}_{-0.5}$ & $3.59^{+1.76}_{-1.41}$$\times$10$^8$  \\
 $r$ & 14.98 $\pm$ 0.002 & $-17.2^{+0.4}_{-0.5}$ & $3.20^{+1.57}_{-1.26}$$\times$10$^8$   \\
 $z$ & 14.63 $\pm$ 0.002 & $-17.5^{+0.4}_{-0.5}$ & $4.42^{+2.17}_{-1.74}$$\times$10$^8$  \\
\hline                                   
\end{tabular}
}
\end{table}

\section{The compact stellar system within the stellar shell}
\label{sec:4}

In this Section we focus on the properties 
of the compact stellar system embedded in the diffuse substructure.
Section~\ref{sec:phot} provides photometric measurements and
Section~\ref{sec:spops} covers stellar population modelling.

\subsection{Photometry}\label{sec:phot}

\begin{figure}
{\centering
	\includegraphics[width=7.5cm]{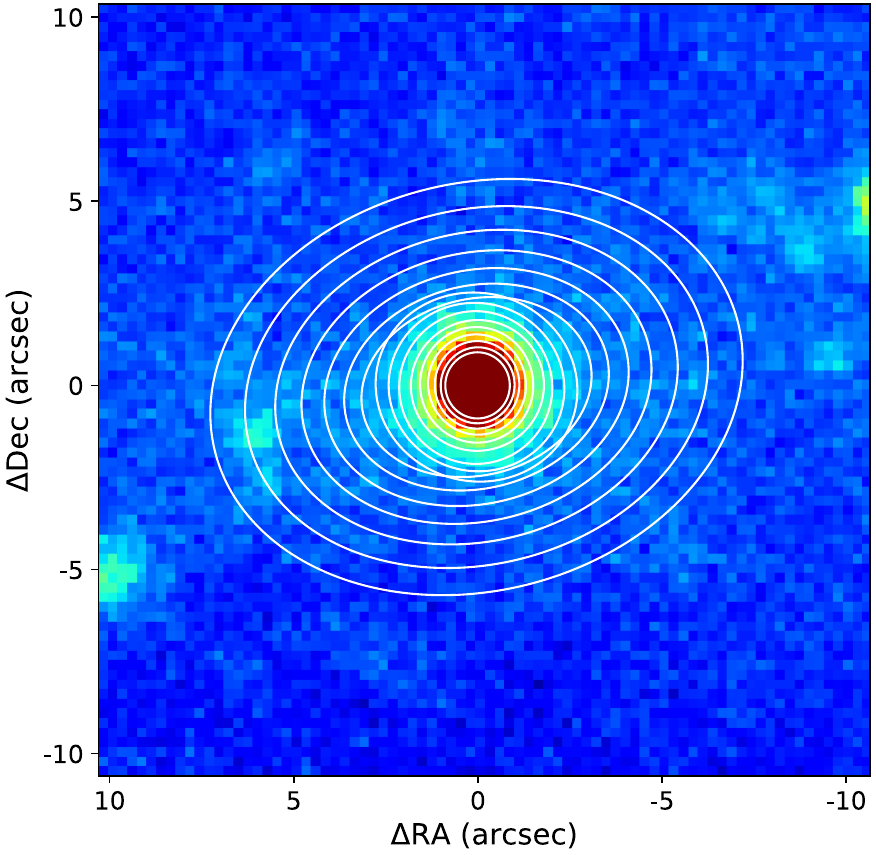}
 }
    \caption{DESI Legacy Survey image of the compact stellar system and surroundings, with elliptical isophote contours overlaid.  North is up and East is left.
    The compact, round star cluster transitions to a flattened structure at $\sim$~2--3~arcsec that likely represents tidal tails (also visible in Figure~\ref{fig-cluster}).
    }
    \label{fig:isophote}
\end{figure}

For photometric analysis of the CSS, we make use of imaging in $griz$ from DESI LS (DR9) and in near-ultraviolet (NUV) from {\it GALEX} (\citealt{galex}).
The CSS is visually slightly extended in the optical imaging compared to surrounding stars, 
and determining its properties in detail requires photometric modeling convolved with a 
point-spread function (PSF).
We use bright stars in the image to create a PSF model using {\tt photutils}
(finding a full width at half maximum of FWHM~$\sim$~1.0--1.5~arcsec), and confirm that the 
radial surface brightness profile of the CSS is more extended than the PSF. 
We also find clear signatures of tidal disruption, with strong changes of position angle, ellipticity and surface brightness over the radial range of $\sim$~2--3~arcsec ($\sim$~200--300 pc) that transition into a larger, flattened, symmetric structure extending East--West
(see Figure~\ref{fig:isophote}).

We fit a two-dimensional (2D) model to the CSS using GALFIT \citep{Peng2010}, 
applied first to the stacked ($r+i+z$) ``detection'' image.
We adopt a \citet{Moffat1969} model for the CSS, convolved with the PSF, with the diffuse background of tidal debris modelled as a 2D Gaussian plus flat plane. 
The total light in the Moffat component does not converge, and we truncate it at an estimated tidal radius of $r_{\rm t} \approx 2.8$~arcsec.
The best-fitting half-light radius $R_{\rm h}$,
axis ratio $b/a$ and position angle are reported in Table~\ref{tab:prop}.
The uncertainties are dominated by variations in the adopted tidal radius.
A key result here is $R_{\rm h} = 0.13\pm0.05$ arcsec, which corresponds to $14\pm5$~pc at a distance of 22.2 Mpc.

After applying GALFIT to the detection image, we measure photometry in separate filters
by fixing all the model parameters except for the brightness normalisation (while using the appropriate PSF for each band).
The total $g$-band magnitude is reported in Table~\ref{tab:prop},
with the uncertainty again derived through varying the tidal truncation.
The colours are insensitive to the truncation radius, and instead we use
Monte Carlo simulations to estimate these uncertainties
(since the uncertainties reported directly by GALFIT are unrealistic).
First, we apply Gaussian perturbations to each image based on the error map.  To account for correlated noise between pixels, we magnify these perturbations by a factor of $\sqrt{\rm NEA}$, 
where the Noise Equivalent Area is 
${\rm NEA}=4\pi\times({\rm FWHM}/2.355)^2$
and the units are pixel$^2$.
Then we repeat the fitting procedure with GALFIT described above. 
We carry out this simulation 500 times per image and obtain a suite of parameter measurements.
We finally adopt the standard deviation of each parameter as an estimate of measurement uncertainty.

The optical colours of the CSS appear to be slightly bluer than those of the surrounding diffuse substructure (see Table~\ref{tab-photometry}),
and will be modelled in the next Section.

\begin{table}
	\centering
        \renewcommand{\arraystretch}{1.2}
	\caption{Properties of the compact stellar system.
 These include right ascension and declination coordinates, projected galactocentric radius $R_{\rm G}$, extinction-corrected $g$-band magnitude (apparent and absolute, $g_0$ and $M_{g,0}$),
 colours in various filter combinations,
 half-light radius $R_{\rm h}$,
 position angle, axis ratio $b/a$,
 stellar mass $M_\star$, mass-weighted age and metallicity [M/H].
 The uncertainties are for a fixed distance.}
	\label{tab:example_table}
	\begin{tabular}{lcc} 
		\hline
		Property & Value & Units\\
		\hline
		RA & 348.6439 & deg \\
		Dec & $-$43.6044 & deg \\
  $R_{\rm G}$ & 2.35 & arcmin \\
  $R_{\rm G}$ & 15.2 & kpc  \\
		$g_0$ & $20.69 \pm 0.07$  & mag \\
   $M_{g,0}$ & $-11.02 \pm 0.07$ & mag  \\
     $({\rm NUV}-g)_0$ & $2.29 \pm 0.60$ & mag\\
 $(g-r)_0$ & $0.54 \pm 0.06$ & mag  \\
 $(g-i)_0$ & $0.72 \pm 0.06$ & mag  \\
 $(g-z)_0$ & $0.82 \pm 0.07$ & mag  \\
  $R_\mathrm{h}$ & $0.13 \pm 0.05$ & arcsec  \\
 $R_\mathrm{h}$ & $14 \pm 5$ & pc  \\
 P.A. & $18.07 \pm 25.73$ & degree\\
 $b/a$ & $0.96 \pm 0.02$\\
 $M_\star$ & $3.7_{-0.7}^{+1.0}\times 10^6$ & $\mathrm{M}_\odot$  \\
 Age & $3.7_{-1.3}^{+2.0}$ & Gyr \\
 $[{\rm M/H}]$ & $+0.13_{-0.03}^{+0.02}$ & dex \\
		\hline
	\end{tabular}\label{tab:prop}
\end{table}

\subsection{Stellar populations}\label{sec:spops}

\begin{figure*}
\centering
\includegraphics[width=1\textwidth]{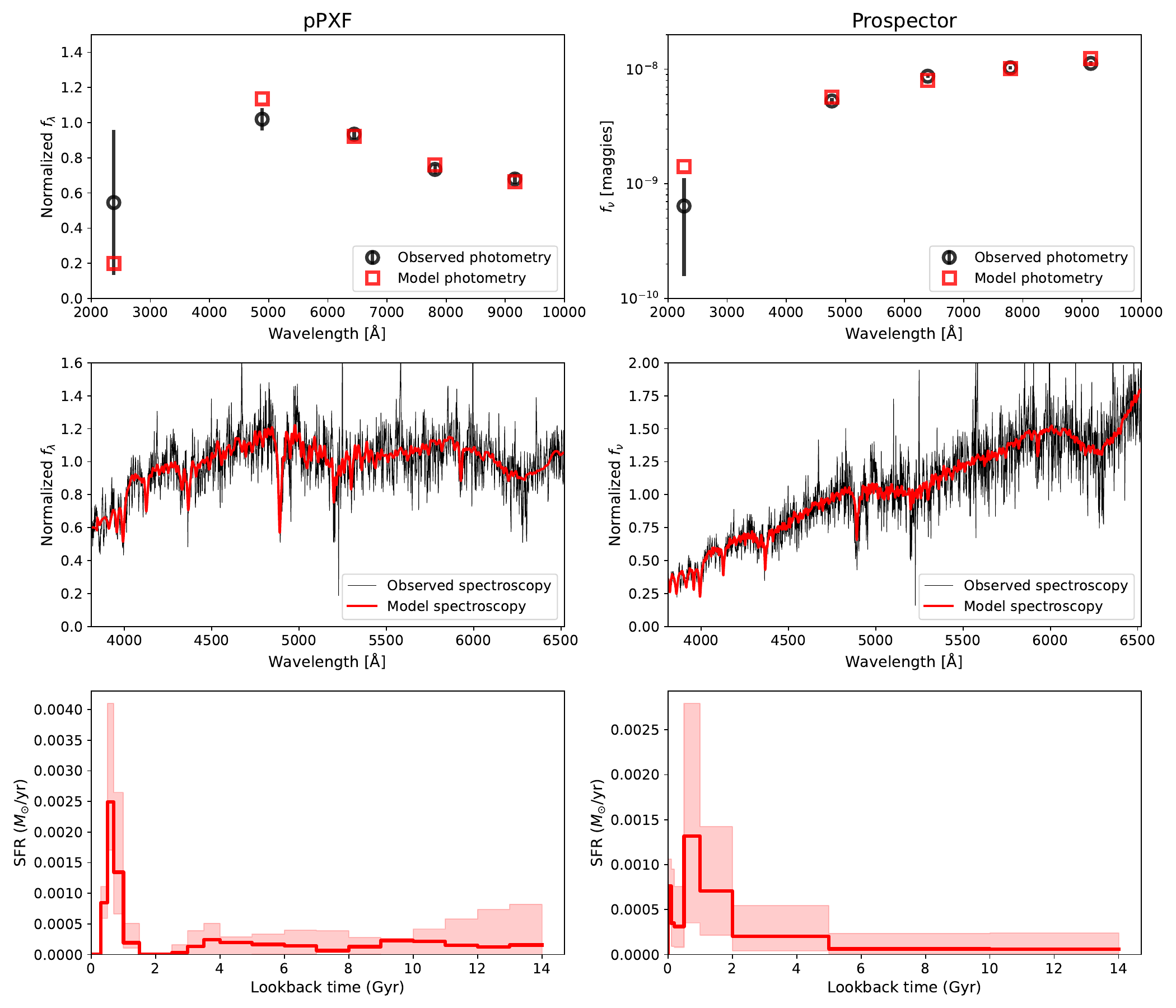}
\caption{Stellar population modelling of the compact stellar system, using independent codes {\tt pPXF} (left column) and Prospector (right column).
The top row shows SED fitting to the photometry,
the middle row shows full spectral fitting
and the bottom row shows the inferred non-parametric star formation rate vs.\ time.
Qualitatively, both codes produce remarkably consistent results, with a very extended star formation history that finished with a strong burst $\sim$~1~Gyr ago.
}
\label{fig:popfit}
\end{figure*}

To model the stellar populations of the CSS, we combine information from the LRIS spectrum (see Section~\ref{sec:data3}) and from the multi-wavelength photometry (Section~\ref{sec:phot}).
Here we use two different modelling codes to test the robustness of the results, shown in Figure \ref{fig:popfit}.
The first is \texttt{pPXF}. For the stellar population models, we use E-MILES single stellar population (SSP) models (\citealt{Vazdekis2016}) with BaSTI library isochrones (\citealt{Hidalgo2018}). These SSP models have 22 different ages from 0.1 to 14 Gyr, and nine different metallicities [M/H] from $-1.79$ to 0.15 dex. 

Initially, we obtain the radial velocity and velocity dispersion by performing a spectrum-only pPXF fitting without regularization. 
Subsequently, we fix the kinematics while running the fitting again for the stellar population with the combination of spectroscopy and photometry. To recover the SFH better, the regularization parameter is set to be 10 according to the spectral S/N. Also, the orders of both additive and multiplicative polynomials are set to 10 to adjust the continuum shape in the fitting, in case of any issue with the flux calibration. pPXF gives a mass-weighted age of $3.84_{-2.06}^{+3.39}$ Gyr, and [M/H] $= +0.11_{-0.02}^{+0.02}$ dex. The uncertainties are obtained through running bootstrapping by resampling the flux residuals.


    The second modelling code is the Bayesian inference code Prospector (\citealt{Johnson2021}), which also allows us to model the photometric and spectroscopic data simultaneously. The stellar models are built based on the MILES stellar spectral library (\citealt{Falcon2011}) and the MIST isochrones (\citealt{Choi2016}) through Flexible Stellar Population Synthesis \citep{Conroy2009}. 

To obtain a non-parametric SFH, we set the time bins to [0, 0.1], [0.1, 0.2], [0.2, 0.5], [0.5, 1.0], [1.0, 2.0], [2.0, 5.0], [5.0, 10.0], [10.0, 14.0] Gyr with the Dirichlet prior. For other parameters, linear uniform priors are adopted, including stellar mass ($6.0<\log M_*/M_{\odot}<7.5$), metallicity ($-2.0<{\rm [M/H]}< +0.2$), and dust ($0<A_V<2.0$). A 20th-order Chebyshev polynomial was employed for calibrating the difference in continuum shapes between photometric and spectroscopic data. We use the dynamic nested sampling algorithm DYNESTY (\citealt{Speagle2020}) to sample the posteriors. From Prospector, we find a mass-weighted age of $3.51_{-1.47}^{+1.93}$ Gyr, and [M/H] of $+0.14_{-0.05}^{+0.03}$ dex, which is consistent with our pPXF numbers mentioned above. Therefore, we adopt the average values of age $= 3.7_{-1.3}^{+2.0}$~Gyr and [M/H] $= +0.13_{-0.03}^{+0.02}$ dex.

Based on the weights of the SSP model at different ages, we recover the non-parametric SFHs from both pPXF and Prospector. The results show remarkable agreement (Figure~\ref{fig:popfit}, lower panels): a relatively constant, low level of star formation from $\sim$~13 to $\sim$~2 Gyr ago, followed by a strong burst $\sim$~1 Gyr ago.

The inferred mass-to-light ratio in the $g$-band is almost the same between pPXF and Prospector, about 1.4. Therefore, our estimate for the CSS stellar mass is $3.7_{-0.7}^{+1.0}\times 10^6$ $\mathrm{M}_\odot$. 

We do not attempt to model the stellar populations of the diffuse substructure, given the lack of either spectroscopy or near-infrared photometry.
However, we can make a qualitative conclusion that it is older than the CSS, given the redder color and implausibility of the metallicity being higher.

\section{Discussion}
\label{sec:discussion}

\subsection{The nature of the compact stellar system}\label{sec:interp}

\begin{figure}
	\includegraphics[width=9.2cm]{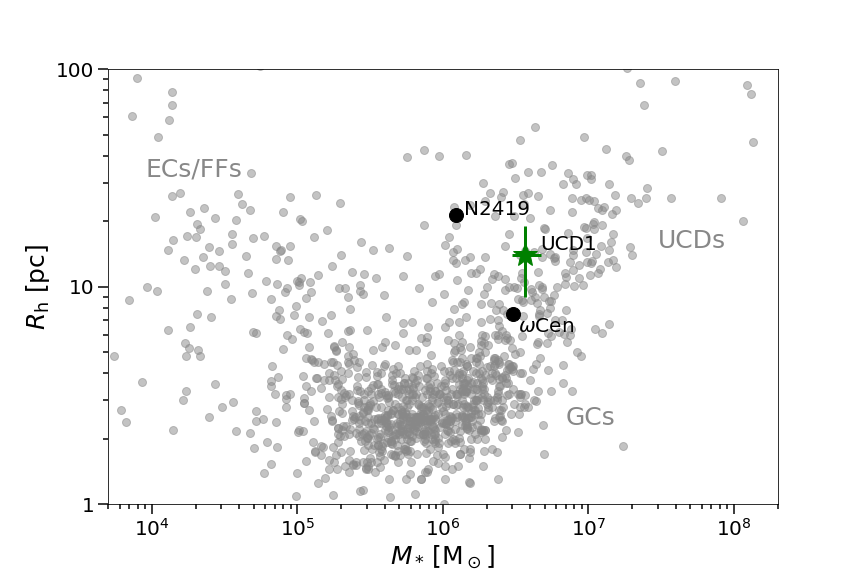}
    \caption{Half-light radius vs. stellar mass for the compact stellar systems. The object lies in the ultra-compact dwarf galaxy (UCD) range, we thereby denote it as UCD1 (green star). For reference, we mark the location of the massive MW star clusters $\omega$~Cen and NGC 2419.  We also denote the approximate locations of globular cluters (GCs), extended clusters (ECs) and faint fuzzies.}
    \label{fig:uber}
\end{figure}

We plot the size and mass of the CSS in Figure~\ref{fig:uber}
along with other objects in the literature for comparison
(see \citealt{Brodie2011}, with updates online\footnote{\url{https://sages.ucolick.org/spectral_database.html}}).
The CSS clearly resides in the parameter space of UCDs, 
i.e.\ larger than classical GCs, and more massive than most of them.
Therefore we dub the object NGC~7531-UCD1, or UCD1 for short -- adding to examples in the literature of ``UCDs in formation'' (see Section~\ref{sec:intro}).
We note that the size and luminosity of UCD1 may well diminish as the tidal stripping progresses, leading to a fainter, more compact system like $\omega$~Cen (Figure~\ref{fig:uber}) -- famous among the GCs of the MW for likely being a remnant of a stripped galaxy (e.g., \citealt{Pfeffer2021}).
The one UCD belonging to the MW, NGC~2419, was originally part of the Sagittarius dwarf, and it is not clear if it was ever an NSC \citep{Pfeffer2021,Davies2024}.

The apparent tidal radius of UCD1 (Figure~\ref{fig:isophote}) allows us to estimate the pericentric distance of its orbit, by assuming that the current isophote disturbances are residuals of a recent pericentric passage rather than an instantaneous response to the local tidal field.
We begin with a standard tidal radius formula of
$r_{\rm t} \simeq r_{\rm G}*(M_\star/3M_{\rm G})^{1/3}$,
where $r_{\rm p}$ is the three-dimensional pericentric distance,
$M_\star$ is the mass of the UCD and $M_{\rm G}$ is the enclosed mass from the host galaxy.  
We assume $M_{\rm G}$ varies linearly with distance, scaled to the circular velocity
from \citet{Pizzella2005}, and solve to find $r_{\rm p} = 11.8^{+2.8}_{-4.2}$~kpc
(with error-bars dominated by the tidal radius uncertainty).
This inference nicely satisfies the consistency criterion that $r_{\rm p} \leq R_{\rm G} =$~15.2 kpc, and suggests that the UCD is fairly close to pericenter.

\subsection{The formation of the tidal debris surrounding NGC~7531-UCD1} \label{sec:sim}

In order to better understand this system, we perform a small suite of $N$-body simulations to determine how a similar morphology can be reproduced. Given the asymmetry of the system, with the bulk of the material sitting on one side of NGC 7531, this is a very unrelaxed structure. As such we choose to model it on radial or nearly radial orbits to determine whether or not we can reproduce the observed morphology. 

Our simulations are performed with \textsc{gadget-3} which is broadly similar to \textsc{gadget-2} \citep{2005MNRAS.364.1105S}. For NGC~7531, we use a scaled-down version of \texttt{MWPotential2014} from \cite{2015ApJS..216...29B}. This scaling is done based on the circular velocity of NGC~7531, 168.6 km~s$^{-1}$ \citep{Pizzella2005} compared to $v_{\rm max} = 225.6$ km~s$^{-1}$ for \texttt{MWPotential2014}. For this scaling, we scale the masses by a factor of $\lambda$ and the distances by a factor of $\lambda^\frac{1}{3}$, in line with the empirically observed virial radius versus size relation \citep{2013ApJ...764L..31K}. We find a value of $\lambda=0.415$ produces a peak circular velocity curve of $168.6$ km~s$^{-1}$. The potential we use is an NFW dark matter halo \citep{1996ApJ...462..563N} with a mass of $3.32\times10^{11} M_\odot$, a scale radius of $11.93$ kpc and a concentration of 15.3. For the disk, we use a Miyamoto--Nagai disk \citep{1975PASJ...27..533M} with a mass of $2.82\times10^{10} M_\odot$, a scale radius of $2.24$ kpc and a scale height of $0.21$ kpc. For the bulge, we use a truncated power law bulge with a mass of $2.08\times10^9 M_\odot$, a power-law slope of $-1.9$ and an exponentially truncated scale radius of 1.42 kpc. 

We model the dwarf that produced the shell with a two-component model which has a \citet{1911MNRAS..71..460P} sphere for the stars and an NFW halo for the dark matter. For the stellar component, we use a stellar mass of $3\times10^8 {\rm M_\odot}$, based on the lower estimate provided in Section~\ref{sec:pholometry2}, and a 3D half-light radius of 1.5 kpc. For the dark matter, we use a virial mass of $M_{200} = 3\times10^{10} {\rm M}_\odot$, a concentration of 11.9 \citep[using the mass--concentration relation in][]{2014MNRAS.441.3359D} and a scale radius of 5.49 kpc. We model each component with $10^6$ particles and a softening of 32.9 pc. 

\begin{figure}
  \includegraphics[width=\columnwidth]{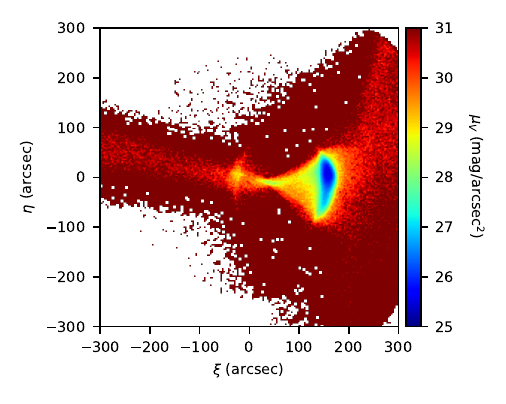}
  \caption{Surface brightness map of the simulated dwarf debris around NGC~7531 which qualitatively matches the observations. This snapshot occurs 60 Myr after the most recent pericentric passage and 1.23 Gyr after the first pericenter.}  
  \label{fig-sim}
\end{figure}

For our near-radial merger, we initialize the system at a distance of 150 kpc along the $z$ axis with a tangential velocity of $2.5$ km~s$^{-1}$ in the $x$ direction. The first pericenter occurs 1.79 Gyr after the simulation begins and produces a shell shortly afterwards. In Figure~\ref{fig-sim} we show a surface brightness map of the simulated dwarf debris $\sim 60$ Myr after the second pericentric passage ($1.23$ Gyr after the first pericentric passage), which has features comparable to those observed. This simulation suggests that two pericentric passages are needed to create the observed shell and the trailing stream. This conclusion matches up well with the recent burst in star formation $\sim$ 1 Gyr ago shown in Figure~\ref{fig:popfit}. We note that 54\% of the debris in Figure~\ref{fig-sim} is fainter than 28 mag/arcsec$^2$, which suggests that the bright shell in NGC 7531 contains almost half of the tidal debris. We also note that this model does not include an NSC and thus does not reproduce the linear feature close to the compact stellar system, i.e. Figure~\ref{fig:isophote}.

\begin{figure*}
  \includegraphics[width=\textwidth]{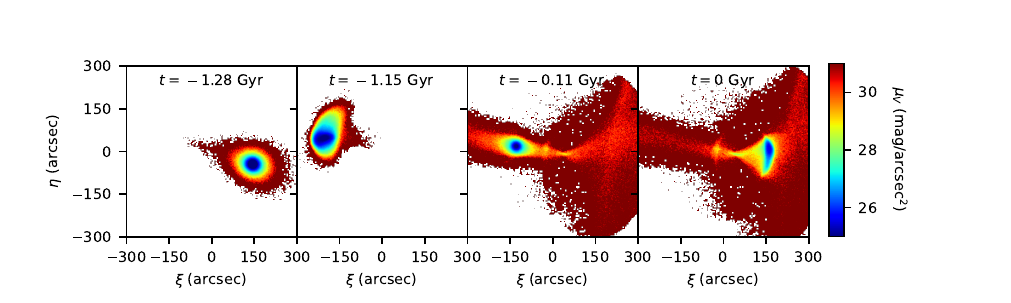}
  \caption{Evolution of the $N$-body simulation which produces a shell similar to what is observed in NGC 7531. From left to right, the panels show the simulated dwarf 50 Myr before the first pericenter ($t=-1.28$ Gyr), 70 Myr after the first pericenter ($t=-1.15$ Gyr), 50 Myr before the second pericenter ($t=-0.11$ Gyr), and 60 Myr after the second pericenter (present day). }  
  \label{fig:sim-panel}
\end{figure*}

In Figure~\ref{fig:sim-panel}, we show the evolution of the dwarf galaxy to create the shell. The first panel shows the dwarf 50 Myr before its first pericenter (1.28 Gyr ago). During this pericentric passage, it becomes significantly distorted and produces an initial shell shown in the second panel, 70 Myr after the pericentric passage. The apocenter occurs 0.66 Gyr ago at a distance of 60.3 kpc, which corresponds to 565 arcsec at the distance of NGC 7531. The third panel shows the dwarf 50 Myr before its second pericenter (0.11 Gyr ago). The final panel shows the structure at the present day with a vibrant shell, a trailing stream on the left, and an additional shell-like low surface brightness structure to the right.
Both of these latter features also appear in the observations (features `c' and `b' in Figure~\ref{fig-NGC7531}, respectively). We note that the shell can be created from just one passage (e.g. second panel of Figure~\ref{fig:sim-panel}), but this would suggest a more recent starburst and would not produce any additional debris.

To better understand how the observed shell was produced, we also simulated the same system with slightly higher and lower angular momenta by giving it a tangential velocity of 0, 5, and 10 km~s$^{-1}$. For the less radial orbits, the resulting debris looks similar but produces a more asymmetric shell, i.e. the overdense region is on one side of the shell. For the radial infall, the overdense region is in the center of the shell but the trailing stream is aligned with the shell, in disagreement with observations. 

We note that our work on NGC~7531 bears some resemblance to previous work on the Umbrella Galaxy NGC~4651 \citep{Foster2014}.
However, in that case the spectroscopy of the star cluster embedded in the stellar stream was not sufficient for inferring its star formation history, which we have shown can provide critical clues to the accretion history.
Also the Umbrella study used a rescaled existing numerical simulation, while for NGC~7531 we have deployed a bespoke simulation for this system.

\section{Conclusions}
\label{sec:conclusions}

We have performed a detailed characterisation of the shell-like tidal debris around NGC~7531 and the CSS embedded within it. We present new, deep images of NGC~7531 taken with an amateur telescope and luminance filter, which enhance the LSB structures around NGC~7531 in unprecedented detail. For the scientific analysis, we have primarily used DESI LS images, Keck/LRIS spectroscopy and tailored N-body simulations. Here, we present our conclusions:

\begin{itemize}
    \item \textit{The compact object embedded in the shell is a UCD associated with NGC 7531:} Our photometric measurements suggest a stellar mass of $3.7_{-0.7}^{+1.0}\times 10^6$ $\mathrm{M}_\odot$ and a half-light radius of $R_{h} = 0.13 \pm 0.05$ arcsec. Our spectroscopic measurements confirm the object is at a similar distance to NGC 753 and, through SED-fitting, we derive an extended SFH with a recent burst of SF. These three factors strongly suggest that the CSS is a UCD, which we hereby name NGC 7531-UCD1.
    
    \item The presence of tidal tails from NGC 7531-UCD1 suggests the system underwent (and is likely ongoing) tidal disruption, adding observational evidence to our knowledge of the NSC--UCD formation pathway and to the pool of known extragalactic examples.

    \item \textit{Our simulations reproduce the formation of the tidal debris:} Our photometric measurements quantify the shell-like debris' mass as $M_\star\sim 3$--$11\times 10^8 M_\odot$, implying a merger mass ratio from $\sim$ 300:1 to 10:1. With this information (and other parameters) we are able to reproduce the merger event, recovering  a near radial orbit of the progenitor dwarf galaxy with two pericentric passages. The timing of the first pericentre passage coincides with the measured star formation enhancement $\sim$~1 Gyr ago.

\end{itemize}

Our detailed study of NGC~7531 and its UCD demonstrates the scientific potential that will be unlocked by next-generation wide-field surveys, such as the Legacy Survey of Space and Time (LSST; \citealt{LSST}), {\it Nancy Grace Roman Space Telescope} (\citealt{roman}), and {\it Euclid} (\citealt{mellier2025}) in resolving star clusters within galactic halos. {\it Euclid}, in particular, has already demonstrated its ability to distinguish GCs across a range of contexts, from within dwarf galaxies to the intracluster field (e.g., \citealt{hunt2025, voggel2025, marleau2025, Saifollahi2025,Romanowsky2025}), and to characterise their properties (e.g., \citealt{Urbano2024,Larsen2025}). With the present study, we have shown that characterising star clusters within tidal debris not only improves our knowledge of star cluster formation and evolution pathways, but the information can also be used to constrain the accretion history of the galaxy and the origin of the debris itself.

\section*{Acknowledgements}

 DMD acknowledges project (PDI2020-114581GB-C21/ AEI / 10.13039/501100011033) and project CNS2022\_136017. JS acknowledges financial support from project PID2022-138896NB-C53 and the Severo Ochoa grant CEX2021-001131-S funded by MCIN/AEI/ 10.13039/501100011033. DMD acknowledges the financial support provided by the Governments of Spain and Arag\'on through their general budgets and  Fondo de Inversiones de Teruel, and the Aragonese Government through the Research Group E16\_23R. SRF acknowledges financial support from the Spanish Ministry of Economy and Competitiveness (MINECO) under grant number AYA2016-75808-R, AYA2017-90589-REDT and S2018/NMT-429, and from the CAM-UCM under grant number PR65/19-22462. SRF acknowledges support from a Spanish postdoctoral fellowship, under grant number 2017-T2/TIC-5592. MAGF acknowledges financial support from the Spanish Ministry of Science and Innovation through the project PID2020-114581GB-C22. JR acknowledges support from the State Research Agency (AEI-MCINN) of the Spanish Ministry of Science and Innovation under the grant "The structure and evolution of galaxies and their central regions" with reference PID2019-105602GB-I00/10.13039/501100011033. JR also acknowledges funding from the University of La Laguna through the Margarita Salas Program from the Spanish Ministry of Universities ref. UNI/551/2021-May 26, and under the EU Next Generation. M.A acknowledges the financial support from the Spanish Ministry of Science and Innovation and the European Union - NextGenerationEU through the Recovery and Resilience Facility project ICTS-MRR-2021-03-CEFCA. AJR was supported by National Science Foundation grant AST-2308390. MAGF acknowledges financial support from the Spanish Ministry of Science and Innovation through the project PID2020-114581GB-C22. SRF acknowledges financial support from the Spanish Ministry of Economy and Competitiveness (MINECO) under grant numbers AYA2016-75808-R, AYA2017-90589-REDT and S2018/NMT-429, and from the CAM-UCM under grant number PR65/19-22462. SRF acknowledges support from a Spanish postdoctoral fellowship, under grant number 2017-T2/TIC-5592. JMC acknowledges the support received from The Leiden Observatory, which has provided facilities and computer infrastructure for carrying out part of this work, and in particular, the support received from Prof. Dr. Konrad Kuijken.
 
This work was partly done using GNU Astronomy Utilities (Gnuastro, ascl.net/1801.009) version 0.17. Work on Gnuastro has been funded by the Japanese MEXT scholarship and its Grant-in-Aid for Scientific Research (21244012, 24253003), the European Research Council (ERC) advanced grant 339659-MUSICOS, and from the Spanish Ministry of Economy and Competitiveness (MINECO) under grant number AYA2016-76219-P.

M.A acknowledges the financial support from the Spanish Ministry of Science and Innovation and the European Union - NextGenerationEU through the Recovery and Resilience Facility project ICTS-MRR-2021-03-CEFCA,

Some of the data presented herein were obtained at Keck Observatory, which is a private 501(c)3 non-profit organization operated as a scientific partnership among the California Institute of Technology, the University of California, and the National Aeronautics and Space Administration. The Observatory was made possible by the generous financial support of the W. M. Keck Foundation. 
The authors wish to recognize and acknowledge the very significant cultural role and reverence that the summit of Maunakea has always had within the Native Hawaiian community. We are most fortunate to have the opportunity to conduct observations from this mountain.

%
\bibliographystyle{aa} 
\bibliography{ref} 
%

\end{document}